\documentclass[prd,a4paper,superscriptaddress,nofootinbib]{revtex4}
\usepackage{amsmath,amssymb,epsfig,natbib,bm}

\begin{document}

\newcommand{\BW}[2]{#2}   


\newcommand{\Mpc}{\text{Mpc}}
\newcommand{\m}{\bar{m}}
\newcommand{\dlnfdlnq}{\frac{\text{d}\ln F}{\text{d} \ln q}}
\newcommand{\I}{I}
\newcommand{\rhat}{\hat{r}}
\newcommand{\half}{{\textstyle \frac{1}{2}}}
\newcommand{\third}{{\textstyle \frac{1}{3}}}
\newcommand{\numfrac}[2]{{\textstyle \frac{#1}{#2}}}
\newcommand{\ra}{\rangle}
\newcommand{\la}{\langle}
\newcommand{\qoe}{\numfrac{q}{\epsilon}}
\newcommand{\dlfdlq}{\frac{d \ln F}{d \ln q}}
\newcommand{\ixb}{{(b)}}
\newcommand{\ixnu}{{(\nu)}}
\newcommand{\ixgam}{{(\gamma)}}
\newcommand{\ixc}{{(c)}}
\newcommand{\lm}{{(lm)}}
\newcommand{\ginv}{{(\text{gi})}}
\newcommand{\upi}{\ubar{\pi}}
\newcommand{\usig}{\ubar{\sigma}}
\newcommand{\vort}{\omega}
\newcommand{\eV}{\,\text{eV}}
\newcommand{\cld}{\nabla}

\newcommand{\cla}{\mathcal{A}}
\newcommand{\clb}{\mathcal{B}}
\newcommand{\clc}{\mathcal{C}}

\newcommand{\cle}{\mathcal{E}}
\newcommand{\clf}{\mathcal{F}}
\newcommand{\clg}{\mathcal{G}}
\newcommand{\clh}{\mathcal{H}}
\newcommand{\cli}{\mathcal{I}}
\newcommand{\clj}{\mathcal{J}}
\newcommand{\clk}{\mathcal{K}}
\newcommand{\cll}{\mathcal{L}}
\newcommand{\clm}{\mathcal{M}}
\newcommand{\cln}{\mathcal{N}}
\newcommand{\clo}{\mathcal{O}}
\newcommand{\clp}{\mathcal{P}}
\newcommand{\clq}{\mathcal{Q}}
\newcommand{\clr}{\mathcal{R}}
\newcommand{\cls}{\mathcal{S}}
\newcommand{\clt}{\mathcal{T}}
\newcommand{\clu}{\mathcal{U}}
\newcommand{\clv}{\mathcal{V}}
\newcommand{\clw}{\mathcal{W}}
\newcommand{\clx}{\mathcal{X}}
\newcommand{\cly}{\mathcal{Y}}
\newcommand{\clz}{\mathcal{Z}}
\renewcommand{\H}{\clh}
\newcommand{\sigthom}{\sigma_T}
\newcommand{\Al}{{A_l}}
\newcommand{\Bl}{{B_l}}
\newcommand{\eAl}{e^\Al}
\newcommand{\ix}{{(i)}}
\newcommand{\ixp}{{(i+1)}}
\renewcommand{\k}{\beta}

\newcommand{\HD}{\mathrm{D}}
\newcommand{\curl}{\mathrm{curl}}
\newcommand{\CMBFAST}{\textsc{CMBFAST}}
\newcommand{\Omtot}{\Omega_{\mathrm{tot}}}
\newcommand{\Omb}{\Omega_{\mathrm{b}}}
\newcommand{\Omc}{\Omega_{\mathrm{c}}}
\newcommand{\Omm}{\Omega_{\mathrm{m}}}
\newcommand{\Oml}{\Omega_\Lambda}
\newcommand{\OmK}{\Omega_K}

\title{Evolution of cosmological dark matter perturbations}

\author{Antony Lewis}
 \email{Antony@AntonyLewis.com}
 \affiliation{DAMTP, CMS, Wilberforce Road, Cambridge CB3 0WA, UK.}

\author{Anthony Challinor}
 \email{A.D.Challinor@mrao.cam.ac.uk}
 \affiliation{Astrophysics Group, Cavendish Laboratory, Madingley Road, Cambridge CB3 OHE, UK.}

\begin{abstract}
\vspace{\baselineskip}
We discuss the propagation of dark matter perturbations with non-zero
velocity dispersion in cosmological models. 
In particular a non-zero massive neutrino component may well have a significant
effect on the matter power spectrum and cosmic microwave background anisotropy.
We present a covariant analysis of the evolution of a dark
matter distribution via a two-dimensional momentum-integrated hierarchy
of multipole equations. 
This can be expanded in the velocity weight to 
provide accurate approximate equations if the matter is
non-relativistic, and we also perform an expansion in the mass to study the
propagation of relativistic matter perturbations.
We suggest an approximation to the exact
hierarchy that can be used to calculate efficiently the effect of the
massive neutrinos on the CMB power spectra.
We implement the corresponding scalar mode equations
numerically 
achieving a considerable reduction in computation time compared with
previous approaches.
\end{abstract}
\maketitle

\section{Introduction}

Current observational evidence suggests that a significant fraction of
the matter density of the universe is in the form of dark matter. Most
of this must have low velocity dispersion to be consistent with large
scale structure, and the cold dark matter (CDM)
model has so far proved to be a good model. However it is quite
possible that the dark matter has non-zero velocity dispersion, in the
form of warm dark matter. There is also good evidence that the
neutrinos have non-zero mass, so there will be a small component of
hot dark matter in the form of massive neutrinos. In this paper we
study the evolution of dark matter perturbations with a non-zero
velocity dispersion, focusing especially on the massive neutrinos and
their effect on the Cosmic Microwave Background (CMB) anisotropy.

One of the useful products of forthcoming high-resolution CMB
observations should be good limits on the massive neutrino masses. 
Atmospheric neutrino oscillation observations currently provide three-sigma evidence
for a mass difference $\Delta m^2 \sim (3\pm 2)\times 10^{-3}
\eV^2$~\cite{Kamiokande01}. This is consistent with a predominantly
$\nu_\mu$-$\nu_\tau$ oscillation though more complicated interpretations are
possible. Solar neutrino 
data show a much smaller mass difference $\Delta m^2 \sim (3\pm
1)\times 10^{-4} \eV^2$~\cite{Barger02}, suggesting that two of the
neutrino mass eigenstates are likely to be nearly degenerate or very small.
The neutrino
mass is related to $\Omega_\nu$, the ratio of the density of massive
neutrinos to the critical density required for a
flat universe, by (see e.g. Ref.~\cite{Peacock99})
\begin{equation}
\Omega_\nu h^2 = \frac{\sum m_\nu}{93.8 \eV}.
\end{equation}
For $h\sim 0.7$ the neutrino oscillations therefore provide
the bound that $\Omega_\nu \agt 10^{-3}$. Recently there has
also been controversial evidence for double beta decay~\cite{Klapdor02,Barger02}, 
which could place interesting lower limits on the neutrino
mass eigenstates if confirmed. Current cosmological evidence is not
very constraining, suggesting only that $\sum m_\nu \alt
4.2\eV$ and hence $\Omega_\nu \alt 0.1$~\cite{Wang01}.

If $\Omega_\nu\sim
10^{-3}$ the observable effect on the cosmology is limited, with the CMB temperature
and polarization power spectra changing by less than a percent
relative to the corresponding pure CDM model. However at slightly
larger values the effects can become important.  For $\Omega_\nu\sim 3\times 10^{-3}$ there are
changes in the electric polarization power spectrum at the few
percent level. For $\Omega_\nu \agt 5\times 10^{-3}$ the effect on the
temperature power spectrum becomes significant at the percent
level. Though it might be more natural for all the neutrino
masses to be very small, three neutrinos of approximately degenerate mass ($\Delta m <
0.07 \eV$) would be consistent with current data and could easily
be cosmologically interesting.

Even if the contribution is small, it is essential to be able to include the effect of
massive neutrinos in the theoretical calculations, both to constrain
the neutrino masses themselves, and for accurate recovery of the other
cosmological parameters. Even if neutrino experiments suggest that the
masses are very small we would also like to check independently that
this is consistent with cosmology.

In this paper we analyse the propagation of dark matter perturbations,
presenting evolution equations that can be used for fast numerical
implementations. We shall focus on linearized massive neutrino perturbations,
though our results could be applied equally well to other forms of
dark or interacting non-massless matter. In particular the low-energy
equations could be used for the efficient propagation of warm dark
matter perturbations. We also provide exact equations that may be a
useful starting point for studying non-linear evolution.

We use a covariant expansion of the dark matter
distribution function following Ref.~\cite{Ellis83}, giving a set of exact multipole equations for the propagation of
collisionless matter.  To compute the matter power spectrum and CMB anisotropies in models
with hot or warm dark matter 
we linearize the multipole equations about an exact
Friedmann-Robertson-Walker (FRW) model. 
 The covariant analysis yields sets of
physically transparent equations that can be split into scalar, vector and
tensor parts as required for numerical implementation (for previous
applications of this approach see Refs.~\cite{Ellis83,Challinor99,Gebbie99,Gebbie99b}).
We derive useful limiting cases when the
matter is highly relativistic and non-relativistic, and also  discuss
an accurate approximate scheme that interpolates between the two
regimes. 
Though we present a covariant analysis here, our approximation techniques
could be applied equally well using metric variables, for example the synchronous or
Newtonian gauge formalism of Ref.~\cite{Ma95}, or the `gauge-ready' formalism of Ref.~\cite{Hwang02}.

For parameter estimation and hypothesis testing with forthcoming CMB
data a large number of accurate theoretical power spectra will need to be
generated, even if fast Monte Carlo sampling techniques are employed.
This can be very
time consuming even using efficient codes such as~\CMBFAST~\cite{Seljak96} or
parallelized derivatives such as CAMB~\cite{Lewis99}. Using
current methods the massive neutrino evolution dominates the
computation time, so our methods for speeding up the neutrino evolution
should be useful. Our numerical code can generate CMB power
spectra (including polarization) two or three times faster than CMBFAST
on one processor, and in under a second on a modern 16
processor machine (non-flat models take about twice as long).

We employ general relativity with signature $(+ ---)$ and use
natural units where $c=1$.


\section{Covariant multipole equations}

We analyse the propagation of the dark matter with respect to a
reference 4-velocity $u^a$. The variables we use are then physically
meaningful and can in principle be measured directly by an observer
moving with this velocity. We decompose the 4-momentum of a particle
of mass $m$ with respect to $u^a$ as
\begin{equation}
p^a=Eu^a+\lambda e^a ,
\end{equation}
where $E$ is the energy, $\lambda =\sqrt{E^2-m^2}$ is the 3-momentum
and $e^a e_a=-1$. We shall assume
that the dark matter is collisionless so that the  distribution
function $f=f(x^a,p^a)$ obeys the Liouville equation 
\begin{equation}
\partial_\tau f  = 0
\end{equation}
along a path described by an affine parameter $\tau$
(defined so that $\partial_\tau x^a = p^a$).

The $e^a$ dependence of the distribution function $f=f(x,E,e^a)$ can
be expanded in terms of irreducible multipoles as
\begin{equation}
f=\sum_{l=0}^\infty F_\Al e^\Al = F + F_a e^a + F_{ab} e^a e^b + \dots,
\end{equation}
where the tensors $F_\Al = F_{\la a_1\dots a_l\ra}$ are projected
(orthogonal to $u^a$)
symmetric and trace-free. The angle brackets denote
the projected symmetric trace free (PSTF) part of the enclosed
indices. This multipole expansion is the covariant equivalent of the usual
spherical harmonic expansion.  Substituting into the Liouville equation we have
\begin{equation}
\sum_{l=0}^\infty \left( \partial_E F_\Al \partial_\tau E\, e^\Al +
p^a \cld_a F_\Al e^\Al + lF_\Al p^a\cld_a e^{a_l} e^{A_{l-1}} \right)=0.
\end{equation}
The propagation is governed by the geodesic equation $p^a \cld_a
p^b=0$, and the component in the $u^a$ direction gives
\begin{equation}
(E/\lambda)\partial_\tau E =
\partial_\tau \lambda = E^2 A_a e^a + E\lambda(\sigma_{ab}e^ae^b - H).
\label{geodesic}
\end{equation}
This then implies that
\begin{equation}
h^a{}_b p^c\cld_c e^b = - \numfrac{E^2}{\lambda}(A^a + e^a A^c e^a) - E(
e^b \vort_b{}^a + \sigma_{bc} e^b e^c e^a +\sigma^a{}_b e^b),
\end{equation}
where $h_{ab}$ is the projection tensor into the
space orthogonal to $u^a$. Here,
the local Hubble parameter, acceleration, shear and vorticity are given by
\begin{equation}
H=\numfrac{1}{3}\cld^a u_a,\quad\quad A_a =\dot{u}_a \equiv u^b \nabla_b
u_a, \quad\quad \sigma_{ab} = \HD_{\la a} u_{b\ra},
\quad\quad\vort_{ab} = \HD_{[a} u_{b]},
\end{equation}
and $\HD_a X_{a_1\dots a_l} = h_a{}^b h_{a_1}{}^{b_1}\dots
h_{a_l}{}^{b_l} \cld_b X_{b_1\dots b_l}$ is the spatial covariant derivative.
Using these relations in the Liouville equation and equating
irreducible PSTF terms one obtains a set of equations for the
evolution of the distribution function multipoles:
\begin{multline}
E \,{}_\bot\!\dot{F}_\Al -\ \lambda^2 H \partial_E F_\Al- \numfrac{l+1}{2l+3} \lambda \HD^a F_{a\Al} + \lambda \HD_{\la a_l}
F_{A_{l-1}\ra}- l E  F_{a\la
A_{l-1}}\vort_{a_l\ra}{}^a\\
+\left[ \lambda E \partial_E F_{\la A_{l-1}} -
(l-1)\numfrac{E^2}{\lambda} F_{\la A_{l-1}}\right] A_{a_l\ra} 
 - \numfrac{l+1}{2l+3}\left[ (l+2)\numfrac{E^2}{\lambda} F_{a\Al} + \lambda
E\partial_E F_{a\Al}\right] A^a \quad\quad\quad\quad\quad\quad\\
-\numfrac{l}{2l+3} \left[ 3E F_{a \la A_{l-1}} +
2\lambda^2 \partial_E F_{a \la A_{l-1}}\right] \sigma_{a_l\ra}{}^a
+ \numfrac{(l+1)(l+2)}{(2l+3)(2l+5)} \left[ (l+3)E
F_{ab\Al} + \lambda^2\partial_E F_{ab\Al}\right]\sigma^{ab}
\\
+\left[\lambda^2\partial_E F_{\la A_{l-2}} -
(l-2)E F_{\la A_{l-2}} \right]\sigma_{ a_{l-1}a_l\ra}=0,
\label{FAl}
\end{multline}
where ${}_\bot\!\dot{F}_\Al = h^{B_l}{}_{A_l} u^a \cld_a F_{B_l}$ is the
projected time derivative.  

The dark matter couples to other matter via gravity according to
general relativity. The relevant quantities for the gravitational
interaction are not the distribution function multipoles but rather
the integrated quantities that make up the stress-energy tensor.
Following Ref.~\cite{Ellis83} we therefore integrate over momentum defining the momentum-integrated multipoles
\begin{equation}
J_\Al^\ix \equiv \frac{4\pi (-2)^l (l!)^2}{(2l+1)!}\int^\infty_0
\text{d}\lambda\, \lambda^2 E F_\Al\left(\frac{\lambda}{E}\right)^{n},
\label{Jdef}
\end{equation}
where $n\equiv l+2i$ is the velocity weight. The numerical factor is introduced so that the energy density, momentum density, anisotropic stress, and
isotropic pressure that make up the
stress-energy tensor are simply
\begin{equation}
\rho = J^{(0)}, \quad\quad q_a = J^{(0)}_a, \quad\quad \pi_{ab} =
J^{(0)}_{ab}, \quad\quad p =
\third J^{(1)}.
\end{equation}
Integrating Eq.~\eqref{FAl} over momentum we obtain the set of exact multipole equations
\begin{multline}
{}_\bot\!\dot{J}^\ix_\Al + H\left[ (3+n)J^\ix_\Al + (1-n)
J^{(i+1)}_\Al\right] + \HD^a J^\ix_{a\Al} - \numfrac{l}{2l+1} \HD_{\la
a_l} J^{(i+1)}_{A_{l-1}\ra}- l J^\ix_{a \la A_{l-1}}\vort_{a_l\ra}{}^a\\
+ \numfrac{l}{2l+1}\left[ (l+n+1)  J^\ix_{\la A_{l-1}} +
 (2-n) J^{(i+1)}_{\la A_{l-1}}\right] A_{a_l\ra}
+\left[ (l-n)  J^{(i-1)}_{a\Al} + (n-2) J^\ix_{a\Al}\right] A^a\\
+ \numfrac{l}{2l+3}\left[ (2n+3) J^\ix_{a
\la A_{l-1}} + (2-2n) J^{(i+1)}_{a \la A_{l-1}}\right]\sigma_{a_l\ra}{}^a
+ \left[ (l-n)J^{(i-1)}_{ab\Al} + (n-1)
J^{(i)}_{ab\Al}\right]\sigma^{ab}\\
+\numfrac{l(l-1)}{4l^2-1} \left[ (n-1)
J^{(i+2)}_{\la A_{l-2}} - (l+n+1) J^{(i+1)}_{\la A_{l-2}}\right]
\sigma_{a_{l-1} a_l\ra} = 0. \label{JAl}
\end{multline}
This exact two-dimensional infinite hierarchy of covariant
equations can be used as a starting point for analysing the
propagation of dark matter with arbitrary velocity dispersion.
In the case of collisional matter the right hand side would contain a
collision term. The $(l,i)$ element of the hierarchy involves moments
with $l$ in the range $(l-2,l+2)$ and $i$ in the range $(i-1,i+2)$.
The coefficients in front of the $i-1$ moments vanish for $i=0$, so the
subset of equations with $i\geq 0$ form a closed system for the $i\geq 0$
moments.


\section{Cosmological dark matter}
\label{sec:cosmo}

The application of the two-dimensional multipole hierarchy we consider here
is to dark matter
perturbations in cosmology. The universe is assumed to be nearly
FRW and we linearize by dropping products of any
quantities that vanish in an exact FRW universe. All the $l>0$
multipoles vanish by isotropy in an exact FRW model and are therefore
first order. The only zero-order quantities are the scale factor,
Hubble parameter and the $l=0$ multipoles. To first order about a FRW
universe the momentum-integrated multipole equations~\eqref{JAl} become
\begin{eqnarray}
\dot{J}^\ix_\Al + H\left[ (1-n)J^{(i+1)}_\Al +
(3+n)J^\ix_\Al\right] - \numfrac{l}{2l+1} \HD_{\la a_l}J^{(i+1)}_{A_{l-1}\ra} +
\HD^a J^\ix_{a\Al}\nonumber \quad\quad\quad\quad\\\quad\quad\quad
+\left[ (1+l+n)J^{(i+l-1)} + (3 - l
-n)J^{(i+l)}\right](\delta_{1l}\,\third A_{a_1} -
\delta_{2l}\,\numfrac{2}{15}\sigma_{a_1a_2}) = 0.
\label{multipoles}
\end{eqnarray}
We characterize perturbations to the $J^\ix$ (which do not vanish in the
FRW limit) by their first-order co-moving spatial gradients
\begin{equation}
\chi^\ix_a \equiv S\HD_a J^\ix.
\end{equation}
Here, $S$ is the scale factor defined generally in a perturbed universe
by integrating $\dot{S}=HS$ with initial conditions chosen to ensure that
$\HD_a S$ is first order.
Taking the spatial gradient of the $l=0$ equations and commuting
derivatives gives the propagation equation
\begin{eqnarray}
\dot{\chi}^\ix_a + \left[ (1-2i)J^{(i+1)} +
(3+2i)J^\ix\right]\dot{h}_a + S\HD_a\HD^b J^\ix_b 
+ H\left[(1-2i)\chi^{(i+1)}_a +(3+2i)\chi^\ix_a\right]
=0,
\label{chi_prop}
\end{eqnarray}
where $h_a\equiv \HD_a S$ is the projected gradient of the scale factor.
Note that
\begin{equation}
\dot{h}_a = S(\HD_a H - H A_a)
\end{equation}
is defined uniquely at linear order once a choice is made for $u^a$,
unlike $h_a$ itself which also depends on the choice of surface on which we
impose $S=\text{constant}$ as an initial condition.

The covariant equations~(\ref{multipoles}) and (\ref{chi_prop})
constitute an infinite two-dimensional hierarchy for
 analysing the propagation of linear dark matter perturbations. For massless
particles $E=\lambda$ and the velocity-weighted moments are identical,
$J^\ix_\Al=J^{(i')}_\Al$. The momentum-integrated equations then reduce to
the usual one-dimensional Boltzmann hierarchy used to propagate 
massless neutrino perturbations~\cite{Challinor99}.

For non-relativistic matter the velocity weight $l+2i$ controls the
 magnitude of the momentum-integrated moments.
If we neglect terms with $l+2i > n_\ast$ the order $n_\ast$ equations become
(for $0 \leq 2i < n_\ast -2$)
\begin{equation}
\dot{J}^\ix_{A_{n_\ast-2i}} + (3+n_\ast) H J^\ix_{A_{n_\ast-2i}} = 0.
\end{equation}
Using $H = \dot{S}/S$ this solves to give
\begin{equation}
J^\ix_{A_{n_\ast-2i}} \propto S^{-3}S^{-n_\ast}.
\label{velocitydens}
\end{equation}
As expected the truncated equations will be accurate to within terms
that decay as an $n_\ast$th-order velocity density. In addition to demanding
that $v_\ast \ll 1$, where $v_\ast$ is the typical particle velocity,
an accurate truncation also requires that the perturbations do not vary too
rapidly in space. The spatial derivative terms in the $n_\ast$th-order
equations can only be neglected if $k v_\ast \ll S H$, i.e.\
the free-streaming scale in an expansion time is much less than the
proper wavelength $S/k$ of the fluctuations.

A hierarchy truncated at $n_\ast=2$ was used in Ref.~\cite{Maartens99}
to study stress
effects in dark matter structure formation. For warm or hot dark matter one
can choose $n_\ast$ as large as required to obtain accurate results.

There is a subtlety associated with this low-velocity-weight truncation scheme
that arises because the $l=1$ momentum-integrated moments are not
frame-invariant in linear theory. Under linear changes in the fiducial
velocity, $u^a \rightarrow \tilde{u}^a = u^a + v^a$, one can show that
the moments transform as
\begin{equation}
J^{(i)}_{A_l} \rightarrow \tilde{J}^{(i)}_{A_l} =
J^{(i)}_{A_l} - \tfrac{1}{3}
\delta_{l1} v_{a_1}[(3+2i)J^{(i)} + (1-2i)J^{(i+1)}].
\label{trans}
\end{equation}
For $l=1$ and $i=0$ this reduces to the well-known result
$\tilde{q}^a = q^a - (\rho+p)v^a$. We see that at $l=1$ the transformation
is not homogeneous in the velocity weight. The form of Eq.~\eqref{trans}
implies that the truncation condition $J^{(i)}_{A_l}=0$ for $l+2i > n_\ast$
is only frame-invariant at linear order for $n_\ast$ odd. 

\section{Massive neutrino perturbations}

We now discuss how to propagate massive neutrino perturbations from
the early universe until the present day, with the aim of implementing
the equations numerically in an efficient way.

Assuming instantaneous neutrino decoupling in the early
universe the zero-order distribution function for massive neutrinos is given by the Fermi-Dirac distribution
\begin{equation}
f(q) \propto \left[ \exp\left({\numfrac{\sqrt{q^2 + m_\nu^2 S_d^2}}{k_B T_d S_d}}\right) + 1 \right]^{-1},
\label{equilibrium}
\end{equation}
where $T_d$ and $S_d$ are the temperature and scale factor at neutrino
decoupling and $q\equiv S\lambda$ is the co-moving momentum. We assume the particles will be highly relativistic at
decoupling so the mass term can be dropped. As the universe expands the
momenta redshift and
the neutrinos evolve to become non-relativistic. 

At hydrogen recombination light neutrinos will still be quite
relativistic, and the higher multipoles in the distribution can be
significant. Usually it is assumed that the neutrinos become non-relativistic
well before recombination, and hence the high multipoles have little
effect. For light species it is necessary to evolve the higher
multipoles to get accurate results, and approximate schemes like that
in Ref.~\cite{Hu98;2}, based on a
low-$l$ truncation, do not give accurate results. For such relatively
light species it is nonetheless still important to take account of the
non-zero mass,
as this can effect the matter power spectrum and CMB anisotropy power spectrum at the level of
several percent (though some of this is trivially accounted for by the
change in the background equation of state).

It is usual to assume that if the neutrino mass eigenstates are significant they
will be approximately degenerate. This is reasonable as the observed
mass squared differences are sufficiently small that taking them into
account would make no observable difference to the cosmology for the
foreseeable future. However
it is possible that at some point in the future observations will
be sufficiently accurate that even small mass splittings may have an
observable effect, in which case the neutrino mass eigenstates may need to
be propagated separately.

\subsection*{The general regime}

When the most of the neutrinos are either highly relativistic or highly
non-relativistic one can use a small mass or small velocity
expansion, as discussed below. However in general a significant fraction of the neutrinos have intermediate
velocities and one needs to propagate the distribution function directly
to get accurate results. As the neutrinos are locally in equilibrium prior
to decoupling, the distribution function evolves from the isotropic
form in Eq.~(\ref{equilibrium}), but with a spatial variation in
the temperature. The efficiency with which free-streaming converts
this spatial variation into an angular variation depends on the particle
velocity (and the wavelength of the fluctuation). For this reason the
anisotropies in the distribution function do not simply inherit the
momentum dependence of the spatial variation of the initial (isotropic)
distribution, unlike the case for massless particles.

In terms of the co-moving momentum $q\equiv S\lambda$ and energy
$\epsilon\equiv SE$ the multipole equations~\eqref{FAl} for the propagation of the distribution
function linearize to give
\begin{equation}
\dot{F}_\Al - \qoe \numfrac{l+1}{2l+3}\HD^c F_{c\Al} + \qoe \HD_{\la
a}F_{A_{l-1}\ra} + \delta_{l1}\left(\numfrac{\epsilon}{q} A_{a_1} +
\numfrac{1}{S}\qoe h_{a_1}\right)q \partial_q F + \delta_{l2}\sigma_{a_1
a_2} q
\partial_q F=0,
\label{qpoles}
\end{equation}
where the spacetime derivatives are taken at constant $q$. Note that
\begin{equation}
\nabla_a F_{A_l} |_\lambda = \nabla_a F_{A_l}|_q +
(\nabla_a S/S) q \partial_q F_{A_l}.
\end{equation}
The $l=1$ equation contains the first order variable combination
\begin{equation}
\clv_a \equiv S \HD_a F + h_a q\partial_q F = S \HD_a F |_\lambda
\end{equation}
that covariantly characterizes perturbations to the isotropic part of 
the distribution function. It integrates to give the co-moving
gradient of the energy density:
\begin{equation}
\chi_a = \chi_a^{(0)} \equiv S\HD_a\rho = \frac{4\pi}{S^4}\int_0^\infty
\text{d}q\, q^2 \epsilon \clv_a.
\end{equation}
Differentiating Eq.~(\ref{qpoles}) for $l=0$ with respect to time and
commuting derivatives we find the propagation equation
\begin{equation}
\dot{\clv}_a = \third\qoe S \HD_a \HD^b F_b +
\dot{h}_a q\partial_q F.
\label{propv}
\end{equation}
This equation, together with the $l>0$ multipole equations, allows one
to propagate the distribution function perturbations directly.  The
synchronous gauge and Newtonian gauge equivalent of these
equations are well known~\cite{Ma95} and are what is usually used to
propagate numerically massive neutrino
perturbations~\cite{Ma95,Seljak96,Dod96}. This is slow computationally as one has to
propagate a hierarchy of multipoles for each co-moving momentum, and
perform numerical integrations to compute the stress-energy tensor
that is relevant for coupling to the other
perturbations. Even for one massive species the numerical evolution of its
distribution function dominates the computation time, and it becomes
proportionately worse if one has to evolve several massive species.

Note that when a given co-moving momentum becomes sufficiently
non-relativistic ($k q/\epsilon \ll SH$ with $k$ co-moving wavenumber)
free-streaming
becomes ineffective at converting spatial structure into angular structure,
and the spatial-derivative terms in Eq.~\eqref{qpoles} become
negligible. In this limit, $F_{A_l}(q)$ for $l>2$ is approximately constant.
When nearly all the particles are non-relativistic, constancy of
$F_{A_l}(q)$ leads directly to the scaling of the momentum-integrated moments
given in Eq.~\eqref{velocitydens}.

\subsection*{The relativistic regime}

In the relativistic regime the bulk of the distribution have momenta
large compared to the rest mass $q \gg m_\nu S$. Defining the
dimensionless mass parameter $\m \equiv
m_\nu /k_B T_d S_d$ one can expand Eq.~\eqref{Jdef} in terms of the small quantity $\m S$  (which is the ratio of mass to the typical proper momentum)
giving, in the FRW limit,
\begin{equation}
\rho \approx \rho_0\left(1+ \frac{5}{7\pi^2} \m^2S^2\right), 
\quad\quad p \approx \frac{\rho_0}{3}\left(1- \frac{5}{7\pi^2} \m^2S^2 \right),
\quad\quad J^{(2)} \approx \rho_0\left(1-\frac{15}{7\pi^2}\m^2 S^2 
\right),
\label{highe}
\end{equation}
where $\rho_0$ is the density the neutrino would have if it were
massless. These expressions are correct to order $(\m S)^2$, though
the expansion does not generalize straightforwardly to higher order. More generally one can find a relation between the
multipoles of different velocity weight in the perturbed universe;
using the definition given in Eq.~\eqref{Jdef} and expanding in $\m
S$ we have
\begin{equation}
J_\Al^{(i+1)} \approx J_\Al^{(i)} -  m^2 S^2 \frac{4\pi (-2)^l (l!)^2}{S^4 (2l+1)!}\int^\infty_0
\text{d}q  \,\epsilon \,F_\Al.
\end{equation}
To this order we can evaluate the last integral assuming the neutrinos
are massless. Starting from a locally isotropic distribution, the momentum
dependence of the $F_{A_l}$ follows that of the $l=1$ and $2$ source terms
in Eq.~\eqref{qpoles}, so that
\begin{equation}
F_{A_l} \propto q \partial_q F
\label{init}
\end{equation}
for $l>0$. Using this and integrating by parts we have
\begin{equation}
J_\Al^{(i+1)} \approx J_\Al^{(i)}\left( 1 - \frac{5}{7\pi^2} \m^2S^2\right)
\approx J^{(i)}_{A_l} \sqrt{\frac{3p}{\rho}}
\label{eq:close}
\end{equation}
for $l> 0$. In a similar manner, using $\clv_a \propto q\partial_q
F$ in the massless limit, we find
\begin{equation}
\chi_a^{(i+1)} \approx \chi_a^{(i)}\left( 1 - \frac{5}{7\pi^2} \m^2 S^2
\right) \approx \chi_a^{(i)} \sqrt{\frac{3p}{\rho}}.
\label{eq:close1}
\end{equation}
To this order in $\m S$, Eqs.~\eqref{eq:close} and~\eqref{eq:close1}
provide closure conditions that allows reduction of the two dimensional
momentum-integrated hierarchy, Eqs.~\eqref{multipoles} and~\eqref{chi_prop},
to a one-dimensional hierarchy for propagating the $J^{(0)}_\Al$ and
$\chi_a$ that are required for coupling to other perturbations. The
approximation is good for $(l+2i)\m^2 S^2 \ll 1$.

\subsection*{The non-relativistic regime}

When the bulk of the neutrinos have $q \ll m_\nu S$ the energy
density and pressure evolve in the FRW limit as
\begin{eqnarray}
\rho &\approx& \frac{180 \rho_0}{7\pi^4}\left( \zeta_3 \m S +
\frac{15\zeta_5}{2 \m S} - \frac{945\zeta_7}{16 (\m S)^3} +\dots\right),
\nonumber
\\
p &\approx& \frac{900\rho_0}{7\pi^4}\left( \frac{\zeta_5}{\m S} -
\frac{63}{4}\frac{\zeta_7}{(\m S)^3} + \dots\right).
\label{lowe}
\end{eqnarray}
The value of $\m$ today ($S=1$) is related to the neutrino mass
by
\begin{equation}
m_\nu \approx 1.68\times 10^{-4} \m \eV.
\end{equation}
For masses of order $0.2\eV$ or larger $ \m \agt 10^3$,
and the mean speed today is less than about $10^{-3}c$. Even in the
low mass case the neutrinos will therefore have been non-relativistic for a
significant fraction of the universe's evolution.

Since the velocities are becoming small one can accurately
truncate the momentum-integrated hierarchy, Eq.~\eqref{multipoles},
at some fairly low velocity
weight. However in order to do this one does need to know initial values
in the non-relativistic era for those momentum-integrated multipoles that are
retained after truncation. Generally, these cannot be obtained from
the one-dimensional hierarchy that is integrated in the relativistic regime
since the eras of applicability of the relativistic and non-relativistic
approximations do not overlap. (However, a way of interpolating between these
two regimes with sufficient accuracy for CMB computations is discussed below.)
One must therefore integrate the distribution function directly,
using equations~\eqref{qpoles} and~\eqref{propv},
until one enters the non-relativistic regime, at which point
one can switch over to a truncated momentum-integrated hierarchy. 

The evolution of small-scale perturbations depends more critically on
the higher velocity weight multipoles due to the derivative coupling
in Eq.~\eqref{multipoles}. 
Performing a mode expansion in wavenumber as in the appendix, the
truncated hierarchy only becomes accurate at rather later times for small wavenumbers, as discussed in Sec.~\ref{sec:cosmo}.

\subsection*{An approximate scheme}

The relic neutrinos themselves have very low energy and will remain
unobservable for the foreseeable future. We are therefore not
interested in the neutrino perturbations {\it per se}, just in their
effect on the matter and photon perturbations. In this subsection we
suggest an efficient approximate scheme, based on results obtained
above, for propagating the massive neutrino perturbations into the
non-relativistic regime which we have found to be sufficiently accurate for the
computation of the CMB power spectra at the one percent level. This
approximate scheme is inadequate for the matter power spectrum, which
is most efficiently calculated using a momentum grid in the relativistic era
matched onto the truncated momentum-integrated hierarchy when the
neutrinos are non-relativistic.

If we use Eqs.~\eqref{highe} and~\eqref{eq:close}, we can write the
momentum-integrated hierarchy, Eqs.~\eqref{multipoles} and~\eqref{chi_prop},
for $i=0$ during the relativistic era in the form
\begin{eqnarray}
\dot{J}^{(0)}_\Al + H\left[ (3+l)-r(l-1)\right]J^{(0)}_\Al -
r\numfrac{l}{2l+1} \HD_{\la a_l}J^{(0)}_{A_{l-1}\ra} +
\HD^a J^{(0)}_{a\Al} + \delta_{1l} \rho A_{a_1} (1+w)
-\delta_{2l}\frac{2 w\rho}{5}(5-3 w)\sigma_{a_1a_2} &=& 0,
\label{crude} \\
\dot{\chi}_a + H(r+3)\chi_a + \HD_a \HD^b J_b^{(0)} + 3\rho(1+w) \dot{h}_a
&=& 0, \label{crude2}
\end{eqnarray}
where $w\equiv p/\rho$ and $r=(3 w)^{1/2}$. These equations can be used
to evolve the perturbations accurately while the majority of the
particles are relativistic. 

To motivate the approximate scheme, we note that
massive neutrinos \emph{perturbations} affect the CMB power spectra
in the acoustic region mainly by the damping effect on the gravitational
potential of their free-streaming on these scales~\cite{Dod96}. (Other effects,
including changes in the position of the acoustic peaks, arise mainly from the variation
in the background equation of state.) The effect on the potential is
insensitive to the late time evolution of small-scale neutrino perturbations,
so these can be computed in a rather coarse manner at later times.

\begin{figure}
\begin{center}
\epsfig{figure=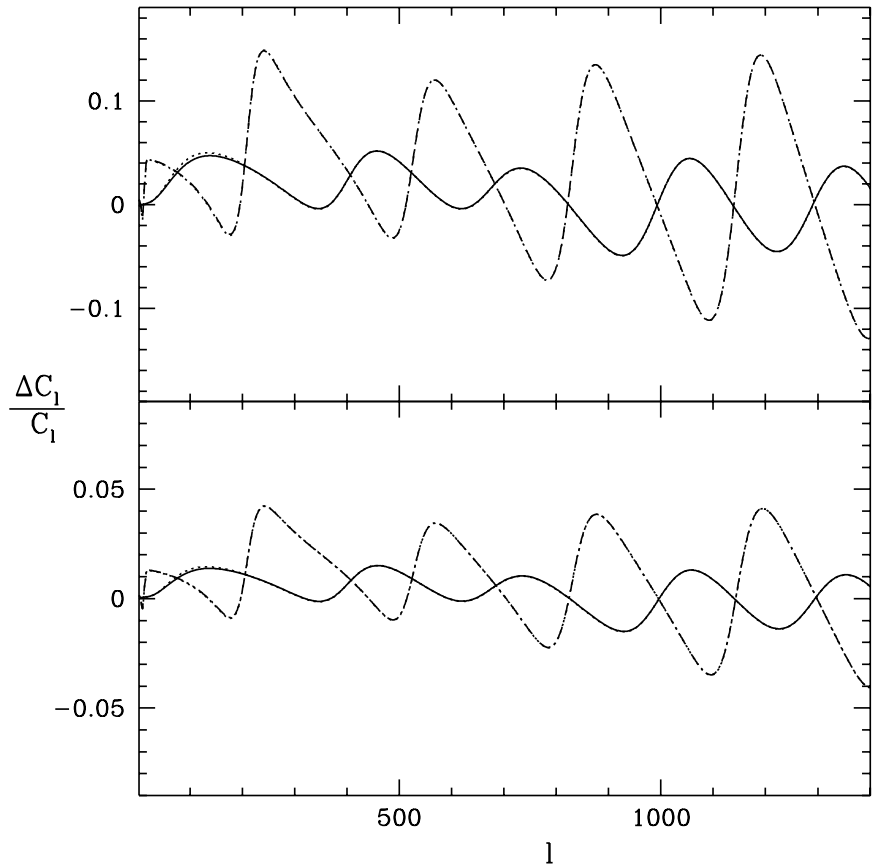,angle=0,width=12.5cm} 
\caption{The change in the scalar CMB temperature (solid lines) and E
polarization (dashed line) power spectra compared to a CDM model, with $\Omega_\nu
h^2 = 6.7\times 10^{-3}$, $N_\nu=3$ (top panel) and $\Omega_\nu
h^2 = 2\times 10^{-3}$, $N_\nu=1$ (bottom panel) computed by integrating the
distribution function perturbations. The dotted lines show the results
from using the approximate Eq.~\eqref{crude}, which agree very closely
almost everywhere. We used $h=0.69$, $\Omega_\Lambda=0.7$,
$\Omega_K=0$, $\Omega_b h^2 = 0.022$. \label{cl_err}}
\end{center}
\end{figure}

At late times the higher velocity weight multipoles fall off rapidly.
Truncating the full momentum-integrated hierarchy at $n_\ast=3$,
one can show that $J_a^{(1)} \approx r J^{(0)}_a$ and $\chi_a^{(1)}
\approx r\chi^{(0)}_a$ are solutions on large scales if $r \approx 5 w$.
Note that these conditions are frame-invariant to the order of our
velocity-weight truncation, and that they give the expected scaling
as the momenta are redshifted since $r \propto 1/S^2$. If we use
$r=5w$ in Eqs.~\eqref{crude} and \eqref{crude2}, and drop terms with velocity
weight $n>3$, we obtain a one-dimensional
hierarchy for $\chi_a$ and the $J^{(0)}_{A_l}$ with $l \leq 3$
that is consistent on large
scales with the full two-dimensional hierarchy
truncated at velocity weight 3. This suggests that if we
evolve Eqs.~\eqref{crude} and \eqref{crude2} with $r$ interpolating between
$(3w)^{1/2}$ and $5w$, we obtain an approximate scheme that
is accurate at early times and gives the correct late time behaviour for
large-scale low-$l$ multipoles.
We found that using
\begin{equation}
r = \left( \frac{5}{3}\right)^{\m S/(\m S+200)} (3w)^{[\m S+2]/[\m S+4]}
\end{equation}
works well and gives CMB power spectra accurate at the one percent
level, as shown in
Fig.~\ref{cl_err}. 
The evolution of the neutrino modes is accurate
until $3w$ becomes significantly less than one, as shown in
Fig.~\ref{evolve}, and is also approximately correct at late times on
large scales. Small scales evolve incorrectly at late times, but this
has no observable consequence for the CMB.

The matter power spectrum is more sensitive to the late-time massive neutrino
evolution than the CMB power spectra.
On scales below the horizon size when
the neutrinos become relativistic, free-streaming prevents them from
clustering in the relativistic era and thus suppresses the growth of
CDM perturbations. Once the neutrinos become non-relativistic they infall
and cluster with the dominant CDM~\cite{Hu98;3}.
The approximate scheme described above is not sufficiently
accurate at late times for computing the matter power spectrum, and the
truncated momentum-integrated hierarchy should be propagated
instead. However the late time matter power spectrum only affects the CMB via second
order effects such as lensing, and the CMB temperature anisotropy
therefore does not depend sensitively on its accuracy.

\begin{figure}
\begin{center}
\BW{\epsfig{figure=evolve.ps,angle=0,width=12.5cm} }{
\epsfig{figure=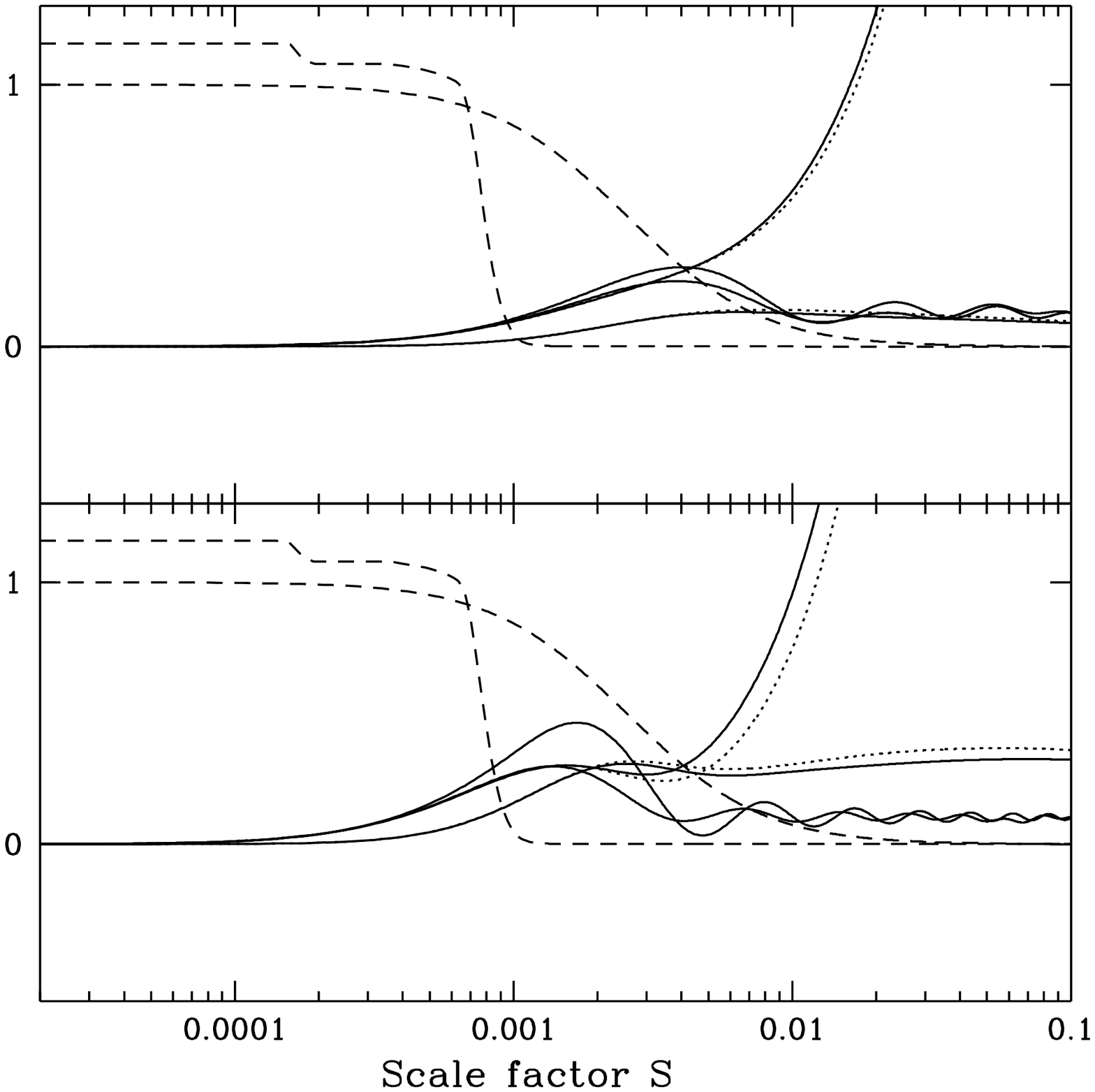,angle=0,width=12.5cm} }

\BW{\caption{The evolution of $k=0.005 \Mpc^{-1}$(top panel) and
$k=0.01\Mpc^{-1}$ (bottom panel) scalar modes, with three degenerate
massive neutrinos giving 
$\Omega_\nu h^2 = 6.7\times 10^{-3}$. The thick dashed and solid lines respectively
show the background ionization fraction and $3 p_\nu/\rho_\nu$. The
solid and dotted lines show the massive neutrino density perturbation and momentum
density,
the faint lines show the result of using Eq.~\eqref{crude}
rather than integrating the distribution function. The short and long dashed lines
show the massless neutrino and photon perturbations. The perturbations
are scaled corresponding to an initial curvature perturbation of $0.2$
and are evaluated in the zero acceleration frame.\label{evolve}}
}{
\caption{The evolution of $k=0.005 \Mpc^{-1}$(top panel) and
$k=0.01\Mpc^{-1}$ (bottom panel) scalar modes, with three degenerate
massive neutrinos giving 
$\Omega_\nu h^2 = 6.7\times 10^{-3}$. The blue and black dashed lines respectively
show the background ionization fraction and $3 p_\nu/\rho_\nu$. The solid black
and red lines show the massive neutrino density perturbation and momentum
density,
the dotted lines show the result of using Eq.~\eqref{crude}
rather than integrating the distribution function. The green and cyan lines
show the massless neutrino and photon perturbations. The perturbations
are scaled corresponding to an initial curvature perturbation of $0.2$
and are evaluated in the zero acceleration frame.\label{evolve}}
}
\end{center}
\end{figure}

\subsection*{Numerical implementation}

The covariant equations can be split into scalar, vector and tensor parts and
then expanded in terms of the appropriate eigenfunctions of the co-moving
Laplacian. This gives sets of equations for the harmonic coefficients that
can be propagated numerically. For details of the scalar and tensor cases
see the appendices.

The background evolution is affected by the neutrino density and
pressure, which is a function only of $\m S$. The numerical integrations
over the distribution function required to compute background quantities
therefore only need be performed
once for each value of $\m S$, and the values can be stored and re-used
with different neutrino masses. Equations~\eqref{highe} and \eqref{lowe}
can be used to calculate quickly the values at low and high values of
$\m S$ respectively.

For the computation of CMB anisotropies we require approximately
one percent accuracy in the power spectrum for
comparison with forthcoming accurate data. We implement the
approximate scheme which is accurate at the one percent level in general,
and much more so if the neutrinos are light. For one massive
species this is about
three times faster than propagating the distribution function
directly, and the massive neutrino evolution no longer dominates the
computation time.

For comparison, and for computing the matter power spectrum, we also propagate
the multipoles of the distribution function
directly. When the momentum has redshifted sufficiently that the neutrinos
are non-relativistic we switch to propagating the truncated momentum-integrated
hierarchy. We keep terms up to $n_\ast=3$, giving four equations to
propagate (the two highest order equations have analytic solutions
$\propto S^{-3}$). The starting values for the momentum-integrated
variables can be computed by
integrating numerically over momentum at the switch-over point. Since
the higher order variables are most important on small scales the
switch-over point needs to be rather later for modes with large
wavenumbers. This is a little slower than the approximate scheme, but
can be made arbitrarily accurate by delaying the switch-over until the
neutrinos have lower velocities. 

Hot dark matter affects the large-scale tensor power spectrum predominantly
through the change in the background equation of state. Tensor modes
interact with the neutrinos only on sub-horizon scales on which they are
decaying away, rapidly becoming negligible compared to the scalar
perturbations.


\section{Conclusion}

We have derived momentum-integrated multipole equations describing the
propagation of dark matter with non-zero velocity dispersion. In the
relativistic and non-relativistic
regimes there are accurate approximations that can be used to evolve
perturbations efficiently. We focussed on massive neutrino
perturbations, though the same techniques could be used for matter
with a different distribution function.
For distributions with intermediate velocities one 
has to evolve the distribution function directly to get accurate results,
though we found that
for computing the CMB power spectrum a fast approximate scheme was
sufficient to take account of the effect of massive neutrinos.
We have implemented the scalar equations numerically and 
our fast parallelized CMB anisotropy code is publicly
available at http://camb.info.


\section*{Acknowledgements}
AL thanks PPARC and the Leverhulme Trust for support. AC is
supported by a PPARC Postdoctoral Fellowship.
\appendix

\section{The scalar equations }

It is convenient to perform a harmonic expansion in terms of
eigenfunctions $Q^k$ of the co-moving Laplacian $S^2 \HD^a \HD_a$,
\begin{equation}
S^2 \HD^a \HD_a Q^k = k^2 Q^k.
\end{equation}
We define scalar coefficients for the harmonic expansion of PSTF tensors
in terms of 
\begin{equation}
Q^k_\Al \equiv (S/k)^l \HD_{\la a_1}\dots \HD_{a_l\ra}Q^k
\end{equation}
as follows:
\begin{equation}
J_\Al^\ix = \rho \sum_k \I^\ix_l Q^k_\Al, \quad\quad \chi_a^\ix=\rho \sum_k
k \I_0^\ix Q_a^k .
\end{equation}
We leave the $k$-dependence of the expansion coefficients implicit. The acceleration, shear and gradient of the scale factor can
be expanded similarly as
\begin{equation}
A_a = \sum_k \numfrac{k}{S} A Q^k_a,\quad\quad
\sigma_{ab} = \sum_k \numfrac{k}{S} \sigma Q^k_{ab},\quad\quad
h_a = \sum_k k h Q^k_a.
\end{equation}
Inserting the harmonic expansion into the multipole equations gives the
propagation equations
\begin{eqnarray}
\I^\ix_l{}' &+&\H\left[(1-n)\I^{(i+1)}_l + (n-3 w^{(1)})\I^\ix_l\right] + 
k\left[ \numfrac{l+1}{2l+1}\k_{l+1} \I^\ix_{l+1} - \numfrac{l}{2l+1} \I^{(i+1)}_{l-1} \right]\nonumber \\
&-&\delta_{l2}\numfrac{2}{5} k\sigma \left[ (3+n)w^{(i+1)}+ (1-n)w^{(i+2)}\right]
+\delta_{l1} kA\left[ (2+n)w^\ix + (2-n)w^{(i+1)}\right] \nonumber\\
&+&\delta_{l0}3 h'\left[(1-n)w^{(i+1)} + (3+n) w^\ix\right] = 0,
\end{eqnarray}
where $3 w^\ix\equiv J^\ix/\rho$, the dash denotes the derivate with respect
to conformal time, $\H = S'/S$ is the conformal Hubble parameter, and
$\k_l \equiv 1 - (l^2-1)K/k^2$ where $K$ is the curvature constant.

The scalar equations  for the distribution function are obtained by expanding the multipoles in terms of scalar harmonics as
\begin{equation}
\clv_a=F \sum_k k F_0 Q_a^k,
\quad\quad F_\Al= F\frac{(2l+1)!}{(-2)^l (l!)^2}\sum_k F_l Q_\Al^k,
\end{equation}
giving the multipole equations
\begin{equation}
F_l' + k \qoe \left(\numfrac{l+1}{2l+1}\k_{l+1}F_{l+1}
- \numfrac{l}{2l+1} F_{l-1}\right)
+\left[\delta_{2l}\numfrac{2}{15} k\sigma - \delta_{1l}\third k \left(
\tfrac{\epsilon}{q} A + \qoe h\right)
-\delta_{0l}h'\right]\dlnfdlnq=0.
\label{massivenuprop}
\end{equation}
Choosing the frame $u^a$ such that $A_a=0$ these equations are equivalent
to the synchronous gauge equations used by other codes~\cite{Ma95,Seljak96}.
Assuming the neutrinos are initially highly relativistic the initial
conditions for $F_l(q)$ are the same as for massless neutrinos:
\begin{equation}
F_l(q) = -\frac{I^{(0)}_l}{4} \dlnfdlnq.
\end{equation}
Later, when the particles are no longer highly relativistic, we can
calculate the momentum-integrated multipoles by integrating over $q$ 
\begin{equation} 
\I^\ix_l = \frac{4\pi}{\rho S^4}\int _0^\infty \text{d}q\,q^2
\epsilon \left(\frac{q}{\epsilon}\right)^{l+2i}F F_l.
\end{equation}
We choose to use the gauge in which $A_a=0$, and keep terms up to
$l+2i=n_\ast=3$. This gives the four energy-integrated scalar equations
\begin{eqnarray}
I^{(0)}_0{}' +  \H\left(\I_0^{(1)} - 3 w^{(1)}I^{(0)}_0\right) + k I^{(0)}_1
+ 3\left(1+w^{(1)}\right) h' &=& 0
\\
\I^{(1)}_0{}' +\H\left(2-3 w^{(1)}\right)I_0^{(1)} + k \I^{(1)}_1 + 15
w^{(1)}h' &=& 0
\\
I^{(0)}_1{}' + \H\left(1-3w^{(1)}\right) I^{(0)}_1 + \third k\left( 2
\k_2 I^{(0)}_2 -  I^{(1)}_0\right) &=& 0
\\
I^{(0)}_2{}' + \H\left(2-3w^{(1)}\right) I^{(0)}_2 + \numfrac{1}{5}k\left(
3\k_3 I^{(0)}_3 - 2 I^{(1)}_1\right) - 2 k w^{(1)} \sigma &=&0,
\end{eqnarray}
where
\begin{equation}
I^{(0)}_3 \propto \rho^{-1} S^{-6}, \quad\quad\quad I_1^{(1)} \propto \rho^{-1}
S^{-6}.
\end{equation}

The density perturbation and momentum density plotted in Fig.~\ref{evolve}
correspond to the values of $I^{(0)}_0$ and $I^{(0)}_1$
in the $A_a=0$ frame. In the non-covariant
approach $I^{(0)}_0$ would just be the synchronous gauge $\delta\rho/\rho$.
\section{The Tensor Equations}

We follow the procedure in Ref.~\cite{Challinor99;2} and expand in terms of
transverse tensor eigenfunctions $Q_{ab}^k$ of the co-moving Laplacian,
\begin{equation}
S^2 \HD^a \HD_a Q_{ab}^k = k^2 Q_{ab}^k.
\end{equation}
We define scalar coefficients for the harmonic expansion of PSTF
tensors with $l\ge 2$ in terms of 
\begin{equation}
Q^k_\Al \equiv (S/k)^{(l-2)} \HD_{\la a_1}\dots
\HD_{a_{l-2}}Q_{a_{l-1} a_l\ra}^k
\end{equation}
as follows:
\begin{equation}
J_\Al^\ix = \rho \sum_k \I^\ix_l Q^k_\Al, \quad\quad \sigma_{ab} = \sum_k
\frac{k}{S} \sigma Q_{ab}^k.
\end{equation}
We leave the $k$-dependence of the expansion coefficients implicit. 

Inserting the harmonic expansion into the multipole equations gives
the $l\ge 2$ propagation equations
\begin{eqnarray}
\I^\ix_l{}' +\H\left[(1-n)\I^{(i+1)}_l + (n-3 w^{(1)})\I^\ix_l\right] +
k\left[ \numfrac{(l+3)(l-1)}{(2l+1)(l+1)}\k_{l+1} \I^\ix_{l+1} - \numfrac{l}{2l+1} \I^{(i+1)}_{l-1} \right]\nonumber \\
-\numfrac{2}{5}k \sigma\left[ (3+n)w^{(i+1)} + (1-n)w^{(i+2)}\right] =0,
\end{eqnarray}
where $\I_1^\ix = 0$ and $\k_l \equiv 1 - (l^2-3)K/k^2$.
The tensor equations for the propagation of the distribution function
follow by expanding
\begin{eqnarray}
 F_\Al= F \frac{(2l+1)!}{(-2)^l (l!)^2}\sum_k F_l Q_\Al^k,
\end{eqnarray}
giving the multipole equations
\begin{equation}
F_l' + k \qoe \left[\numfrac{(l+3)(l-1)}{(2l+1)(l+1)}\k_{l+1}F_{l+1}
- \numfrac{l}{2l+1} F_{l-1}\right]
+\numfrac{2}{15} k\sigma \dlnfdlnq=0,
\end{equation}
where $F_1 = 0$.



\end{document}